\documentclass[aps,prl,twocolumn,superscriptaddress,showpacs]{revtex4} % UTOBS
\usepackage{graphicx}% Include figure files
\usepackage{dcolumn}% Align table columns on decimal point
\usepackage{bm}% bold math
\usepackage{lineno}
\usepackage{setspace}
\usepackage{amsfonts}
\usepackage{amssymb,amsmath}
\usepackage{xspace}
\usepackage{color}
\usepackage{subfigure}
\newcommand{\ppbar}  {\ensuremath{p\bar{p}}\xspace}

\newcommand{\ttbar}  {\ensuremath{t\bar{t}}\xspace}

\newcommand{\met}{\mbox{${\not\!\!E_T}$}}

\newcommand{\unit}[1]{\ensuremath{\mathrm{\:#1}}}

\def\ttbar {\ensuremath{t\overline t}}
\def\ppbar {\ensuremath{p\overline p}}
\def\Cts   {\ensuremath{\cos \theta^{*}}}
\def\CtsSq   {\ensuremath{\cos^{2} \theta^{*}}}

\begin{document}

\title{Measurement of $W$-Boson Polarization in Top-quark Decay in $\ppbar$ Collisions at $\sqrt{s}$ = 1.96\unit{TeV}}

\affiliation{Institute of Physics, Academia Sinica, Taipei, Taiwan 11529, Republic of China} 
\affiliation{Argonne National Laboratory, Argonne, Illinois 60439, USA} 
\affiliation{University of Athens, 157 71 Athens, Greece} 
\affiliation{Institut de Fisica d'Altes Energies, Universitat Autonoma de Barcelona, E-08193, Bellaterra (Barcelona), Spain} 
\affiliation{Baylor University, Waco, Texas 76798, USA} 
\affiliation{Istituto Nazionale di Fisica Nucleare Bologna, $^{dd}$University of Bologna, I-40127 Bologna, Italy} 
\affiliation{Brandeis University, Waltham, Massachusetts 02254, USA} 
\affiliation{University of California, Davis, Davis, California 95616, USA} 
\affiliation{University of California, Los Angeles, Los Angeles, California 90024, USA} 
\affiliation{University of California, San Diego, La Jolla, California 92093, USA} 
\affiliation{University of California, Santa Barbara, Santa Barbara, California 93106, USA} 
\affiliation{Instituto de Fisica de Cantabria, CSIC-University of Cantabria, 39005 Santander, Spain} 
\affiliation{Carnegie Mellon University, Pittsburgh, Pennsylvania 15213, USA} 
\affiliation{Enrico Fermi Institute, University of Chicago, Chicago, Illinois 60637, USA}
\affiliation{Comenius University, 842 48 Bratislava, Slovakia; Institute of Experimental Physics, 040 01 Kosice, Slovakia} 
\affiliation{Joint Institute for Nuclear Research, RU-141980 Dubna, Russia} 
\affiliation{Duke University, Durham, North Carolina 27708, USA} 
\affiliation{Fermi National Accelerator Laboratory, Batavia, Illinois 60510, USA} 
\affiliation{University of Florida, Gainesville, Florida 32611, USA} 
\affiliation{Laboratori Nazionali di Frascati, Istituto Nazionale di Fisica Nucleare, I-00044 Frascati, Italy} 
\affiliation{University of Geneva, CH-1211 Geneva 4, Switzerland} 
\affiliation{Glasgow University, Glasgow G12 8QQ, United Kingdom} 
\affiliation{Harvard University, Cambridge, Massachusetts 02138, USA} 
\affiliation{Division of High Energy Physics, Department of Physics, University of Helsinki and Helsinki Institute of Physics, FIN-00014, Helsinki, Finland} 
\affiliation{University of Illinois, Urbana, Illinois 61801, USA} 
\affiliation{The Johns Hopkins University, Baltimore, Maryland 21218, USA} 
\affiliation{Institut f\"{u}r Experimentelle Kernphysik, Karlsruhe Institute of Technology, D-76131 Karlsruhe, Germany} 
\affiliation{Center for High Energy Physics: Kyungpook National University, Daegu 702-701, Korea; Seoul National University, Seoul 151-742, Korea; Sungkyunkwan University, Suwon 440-746, Korea; Korea Institute of Science and Technology Information, Daejeon 305-806, Korea; Chonnam National University, Gwangju 500-757, Korea; Chonbuk National University, Jeonju 561-756, Korea} 
\affiliation{Ernest Orlando Lawrence Berkeley National Laboratory, Berkeley, California 94720, USA} 
\affiliation{University of Liverpool, Liverpool L69 7ZE, United Kingdom} 
\affiliation{University College London, London WC1E 6BT, United Kingdom} 
\affiliation{Centro de Investigaciones Energeticas Medioambientales y Tecnologicas, E-28040 Madrid, Spain} 
\affiliation{Massachusetts Institute of Technology, Cambridge, Massachusetts 02139, USA} 
\affiliation{Institute of Particle Physics: McGill University, Montr\'{e}al, Qu\'{e}bec, Canada H3A~2T8; Simon Fraser University, Burnaby, British Columbia, Canada V5A~1S6; University of Toronto, Toronto, Ontario, Canada M5S~1A7; and TRIUMF, Vancouver, British Columbia, Canada V6T~2A3} 
\affiliation{University of Michigan, Ann Arbor, Michigan 48109, USA} 
\affiliation{Michigan State University, East Lansing, Michigan 48824, USA}
\affiliation{Institution for Theoretical and Experimental Physics, ITEP, Moscow 117259, Russia} 
\affiliation{University of New Mexico, Albuquerque, New Mexico 87131, USA} 
\affiliation{Northwestern University, Evanston, Illinois 60208, USA} 
\affiliation{The Ohio State University, Columbus, Ohio 43210, USA} 
\affiliation{Okayama University, Okayama 700-8530, Japan} 
\affiliation{Osaka City University, Osaka 588, Japan} 
\affiliation{University of Oxford, Oxford OX1 3RH, United Kingdom} 
\affiliation{Istituto Nazionale di Fisica Nucleare, Sezione di Padova-Trento, $^{ee}$University of Padova, I-35131 Padova, Italy} 
\affiliation{LPNHE, Universite Pierre et Marie Curie/IN2P3-CNRS, UMR7585, Paris, F-75252 France} 
\affiliation{University of Pennsylvania, Philadelphia, Pennsylvania 19104, USA}
\affiliation{Istituto Nazionale di Fisica Nucleare Pisa, $^{ff}$University of Pisa, $^{gg}$University of Siena and $^{hh}$Scuola Normale Superiore, I-56127 Pisa, Italy} 
\affiliation{University of Pittsburgh, Pittsburgh, Pennsylvania 15260, USA} 
\affiliation{Purdue University, West Lafayette, Indiana 47907, USA} 
\affiliation{University of Rochester, Rochester, New York 14627, USA} 
\affiliation{The Rockefeller University, New York, New York 10021, USA} 
\affiliation{Istituto Nazionale di Fisica Nucleare, Sezione di Roma 1, $^{ii}$Sapienza Universit\`{a} di Roma, I-00185 Roma, Italy} 

\affiliation{Rutgers University, Piscataway, New Jersey 08855, USA} 
\affiliation{Texas A\&M University, College Station, Texas 77843, USA} 
\affiliation{Istituto Nazionale di Fisica Nucleare Trieste/Udine, I-34100 Trieste, $^{jj}$University of Trieste/Udine, I-33100 Udine, Italy} 
\affiliation{University of Tsukuba, Tsukuba, Ibaraki 305, Japan} 
\affiliation{Tufts University, Medford, Massachusetts 02155, USA} 
\affiliation{Waseda University, Tokyo 169, Japan} 
\affiliation{Wayne State University, Detroit, Michigan 48201, USA} 
\affiliation{University of Wisconsin, Madison, Wisconsin 53706, USA} 
\affiliation{Yale University, New Haven, Connecticut 06520, USA} 
\author{T.~Aaltonen}
\affiliation{Division of High Energy Physics, Department of Physics, University of Helsinki and Helsinki Institute of Physics, FIN-00014, Helsinki, Finland}
\author{J.~Adelman}
\affiliation{Enrico Fermi Institute, University of Chicago, Chicago, Illinois 60637, USA}
\author{B.~\'{A}lvarez~Gonz\'{a}lez$^w$}
\affiliation{Instituto de Fisica de Cantabria, CSIC-University of Cantabria, 39005 Santander, Spain}
\author{S.~Amerio$^{ee}$}
\affiliation{Istituto Nazionale di Fisica Nucleare, Sezione di Padova-Trento, $^{ee}$University of Padova, I-35131 Padova, Italy} 

\author{D.~Amidei}
\affiliation{University of Michigan, Ann Arbor, Michigan 48109, USA}
\author{A.~Anastassov}
\affiliation{Northwestern University, Evanston, Illinois 60208, USA}
\author{A.~Annovi}
\affiliation{Laboratori Nazionali di Frascati, Istituto Nazionale di Fisica Nucleare, I-00044 Frascati, Italy}
\author{J.~Antos}
\affiliation{Comenius University, 842 48 Bratislava, Slovakia; Institute of Experimental Physics, 040 01 Kosice, Slovakia}
\author{G.~Apollinari}
\affiliation{Fermi National Accelerator Laboratory, Batavia, Illinois 60510, USA}
\author{J.~Appel}
\affiliation{Fermi National Accelerator Laboratory, Batavia, Illinois 60510, USA}
% ...

\collaboration{CDF Collaboration\footnote{With visitors from $^a$University of Massachusetts Amherst, Amherst, Massachusetts 01003,
$^b$Universiteit Antwerpen, B-2610 Antwerp, Belgium, 
$^c$University of Bristol, Bristol BS8 1TL, United Kingdom,
$^d$Chinese Academy of Sciences, Beijing 100864, China, 
$^e$Istituto Nazionale di Fisica Nucleare, Sezione di Cagliari, 09042 Monserrato (Cagliari), Italy,
$^f$University of California Irvine, Irvine, CA  92697, 
$^g$University of California Santa Cruz, Santa Cruz, CA  95064, 
$^h$Cornell University, Ithaca, NY  14853, 
$^i$University of Cyprus, Nicosia CY-1678, Cyprus, 
$^j$University College Dublin, Dublin 4, Ireland,
$^k$University of Edinburgh, Edinburgh EH9 3JZ, United Kingdom, 
$^l$University of Fukui, Fukui City, Fukui Prefecture, Japan 910-0017,
$^m$Kinki University, Higashi-Osaka City, Japan 577-8502,
$^n$Universidad Iberoamericana, Mexico D.F., Mexico,
$^o$University of Iowa, Iowa City, IA  52242,
$^p$Kansas State University, Manhattan, KS 66506,
$^q$Queen Mary, University of London, London, E1 4NS, England,
$^r$University of Manchester, Manchester M13 9PL, England,
$^s$Muons, Inc., Batavia, IL 60510, 
$^t$Nagasaki Institute of Applied Science, Nagasaki, Japan, 
$^u$University of Notre Dame, Notre Dame, IN 46556,
$^v$Obninsk State University, Obninsk, Russia,
$^w$University de Oviedo, E-33007 Oviedo, Spain, 
$^x$Texas Tech University, Lubbock, TX  79609, 
$^y$IFIC(CSIC-Universitat de Valencia), 56071 Valencia, Spain,
$^z$Universidad Tecnica Federico Santa Maria, 110v Valparaiso, Chile,
$^{aa}$University of Virginia, Charlottesville, VA  22906,
$^{bb}$Bergische Universit\"at Wuppertal, 42097 Wuppertal, Germany,
$^{cc}$Yarmouk University, Irbid 211-63, Jordan,
$^{kk}$On leave from J.~Stefan Institute, Ljubljana, Slovenia, 
}}
\noaffiliation

\date{\today}

\begin{abstract}
We report measurements of the polarization of $W$ bosons from top-quark decays using
2.7~$\rm{fb}^{-1}$ of $p \bar{p}$ collisions collected by the CDF II detector.
Assuming a top-quark mass of 175~GeV/$\rm{c}^2$, three measurements are 
performed. A simultaneous measurement of the fraction of longitudinal ($f_0$) and
right-handed ($f_+$) $W$ bosons yields the model-independent results
$f_{0} = 0.88  \pm 0.11 (\rm{stat}) \pm 0.06 (\rm{syst})$  and
$f_{+} = -0.15 \pm 0.07 (\rm{stat}) \pm 0.06 (\rm{syst})$
with a correlation coefficient of -0.59.
A measurement of $f_0$ ($f_+$) constraining $f_+$ ($f_0$) to its standard model value of $0.0$ (0.7) yields
$f_{0} =  0.70 \pm 0.07 (\rm{stat}) \pm 0.04 (\rm{syst})$ ($f_{+} = -0.01 \pm 0.02 (\rm{stat}) \pm 0.05 (\rm{syst})$).
All these results are consistent with standard model expectations.
\end{abstract}

% taken from http://authors.aps.org/eprint/gateway/pacslist
\pacs{14.65.Ha, 14.80.Cp, 13.85.Ni, 13.85.Qk, 14.70.Dj}

\keywords{Top, Properties, Polarization, Helicity}

\maketitle

%%%%%%%%%%%%%%%% Introduction %%%%%%%%%%%%%%%%%%%%%%%
%%%%%%%%%%%%%%%%%%%%%%%%%%%%%%%%%%%%%%%%%%%%%%%%%%%%%

The top quark is the most massive fundamental particle observed by
experiment~\cite{TopDiscovery}. Due to its large mass, 
in the standard model (SM) the top quark decays before
forming a bound state via the charged current weak interaction into a $W^+$ boson
and a $b$ quark~\cite{ChargeConjugate}, with a branching fraction above 99\%~\cite{PDG2008}.  
This provides a unique opportunity to study the
properties of a ``bare'' quark.  In particular, the $V-A$ structure of the
weak interaction can be tested by reconstructing the polarization of
the $W^+$ boson from top-quark decay. In the standard model (SM)
at tree level~\cite{WtbCoupling}, the $W^+$
boson is expected to have longitudinal polarization $f_{0} = 0.703$,
left-handed polarization $f_- = 0.297$, and right-handed polarization
$f_+ = 3.6 \times 10^{-4}$ for a top-quark mass $m_t = 175$ GeV/$\rm{c}^2$,
a $W$-boson mass $M_W = 80.413$ GeV/$\rm{c}^2$~\cite{WmassCDFrunII}, and a $b$-quark
mass $m_b = 4.79$ GeV/$\rm{c}^2$~\cite{PDG2008}. 
In the limit of $m_b \rightarrow 0$, $f_0 = m_t^2 / (2m_W^2 + m_t^2)$ and $f_+ = 0$.
The higher-order QCD and electroweak radiative corrections
modify these predictions at the 1-2\% (relative) level~\cite{QCDandEWcorr}. 
In beyond-the-SM scenarios, significant deviations from the SM expectation are
possible due to the presence of anomalous couplings~\cite{WtbCoupling} in the $tWb$ vertex.
Measurements of $W$-boson polarization and single top-quark production
together set constraints on the anomalous coupling vector and tensor
form factors~\cite{D0anomelousCoupling}.

In this Letter, we measure the polarization of the $W$ boson from top-quark decay.  
We assume the $\ttbar$ production mechanism is in agreement
with the SM, and we study a data sample enriched in 
$\ttbar \to W^+ b W^- \bar{b} \to \ell \nu b q \bar{q}' \bar{b}$
events where one of the $W$ bosons decays hadronically
and the other leptonically. We apply a likelihood technique based
on the theoretical matrix elements for 
both the dominant signal process, $q\bar{q} \rightarrow \ttbar$, and the
main background process, inclusive production of $W$+jets.
This technique was first
developed for the measurements of top-quark mass and 
$f_0$, constraining $f_+$ to its SM value~\cite{ME}, and 
utilizes the kinematic and topological information from the 
event through integrations over poorly known parton-level quantities.
We express the matrix element in terms of the $W$-boson polarization fractions and the cosine of
the angle $\theta^*$ between the momentum of the charged lepton or down-type quark
in the $W$-boson rest frame and the momentum of the $W$ boson in the top-quark rest frame.
Therefore we extract information on the $W$-boson polarization from both the leptonic and hadronic $W$-boson decays.  
Previous CDF measurements\cite{CDFLatestWHelResults,otherCDFWHelResults} used only information from the leptonic decay 
via the transverse momentum $p_T$~\cite{met} of the lepton, $\Cts$  for the leptonically decaying $W$ boson, 
or the invariant mass of each lepton and $b$-jet pair, which is a function of $\Cts$.  While the information from 
the hadronic $W$-boson decay carries a sign ambiguity in $\Cts$ since we are unable to identify the down-type quark 
jet its inclusion still improves the sensitivity to the $f_0$ polarization fraction.  
The analysis described in this 
letter improves the statistical sensitivity on $f_0$ by 20\%
relative to the best previous CDF measurement~\cite{CDFLatestWHelResults} for the same event sample. 
The latest D0 measurement also utilizes information from both the 
leptonic and hadronic $W$-boson decays using a likelihood fit to the $\Cts$ distributions~\cite{D0WHelResults}.

We report measurements of the $W$-boson polarization for three different
hypotheses of top-quark decay:
(1) model-independent with simultaneous
measurement of $f_0$ and $f_+$; 
(2) anomalous tensor couplings
with measurement of $f_0$ for fixed $f_+$=0; and (3) anomalous right-handed
couplings with measurement of $f_+$ for fixed $f_0=0.70$.

The polarization fractions are determined by maximizing the likelihood $L$
with respect to $f_0$, $f_+$, and the fraction of events consistent with the
$\ttbar$ signal hypothesis, $C_s$,

\begin{equation*}
\hspace*{-0.65mm}
  L(f_0, f_+, C_s) = \prod_{i=1}^{N} C_{s} \frac{P_{s}(x;f_0, f_+)}{\langle A_s(x;f_0, f_+) \rangle} + (1 - C_s) \frac{P_{b}(x)}{\langle A_b(x) \rangle }.
\label{eq01} 
\end{equation*}

\noindent Here $N$ is the number of observed events, and 
$\langle A_s \rangle$ and $\langle A_b \rangle$ refer to the average 
acceptances for $\ttbar$ and $W$+jets background events, respectively.
The dependence of the signal acceptance on the polarization fractions is included
in $\langle A_s \rangle$.
The signal probability $P_{s}$ and background probability $P_{b}$
densities are constructed as in~\cite{MEAT} by integrating over the appropriate
parton-level differential cross section, $d\sigma(y)/dy$,
convolved with proton parton distribution functions ($f(q_1)$ and $f(q_2)$) and detector resolution
effects relating a set of observed variables $x$ to corresponding parton-level quantities $y$, $W({x},{y})$:

\begin{equation*}
\hspace*{-0.65mm}
P({x})  = \sum_{\ perm.} \int  \frac{d\sigma(y)}{dy} f(q_1) f(q_2) dq_1 dq_2 W({x},{y})dy.
\label{eq:prob} 
\end{equation*}

\noindent Here partons are identified with the four highest transverse energy $E_T$~\cite{met} jets, 
and all the corresponding jet-parton permutations are considered.

The signal differential cross section
uses the leading-order matrix element
of the $q\bar q\rightarrow t\bar t$ process~\cite{mahlonparke},
expressed in terms of $\Cts$ and polarization fractions:

\begin{equation*}
  |M|^2 = \frac{g_s^4}{9} F_{lep} \bar{F}_{had} (2 - \beta^2 \mathrm{sin}^2\theta_{qt}) \mathrm{,}
  \label{eqn:ME} \nonumber
\end{equation*}

\noindent where $g_s$ is the strong coupling constant and $\theta_{qt}$
describes the angle between the incoming parton and the top quark in
the rest frame of the incoming partons, and $\beta$ is the speed
of the top quarks in the same rest frame.
The factors $F_{lep}$ and $\bar{F}_{had}$ correspond to the top quarks with a leptonic
and a hadronic $W$-boson decay, such that:

\begin{eqnarray*}
  \nonumber
  F_{lep} & = &  \frac{2 \pi}{3} \frac{g_W^4 m_{\bar{\ell}\nu}^2 }{m_t \Gamma_t}
      ( 2 E_{b}^{*2} + 3 E^*_{b} m_{\bar{\ell}\nu} + m_{b}^2 ) \\
    &   &  ( f_{+} \cdot \frac{3}{8} ( 1 + \Cts )^2  + f_{0} \cdot \frac{3}{4} ( 1 - \CtsSq ) \\
    &   &  +  (1-f_0-f_+) \cdot \frac{3}{8} (1- \Cts)^2 ).
  \label{eqn:lepFCts} \\ \nonumber
\end{eqnarray*}

\noindent Here $g_W$ is the weak coupling constant,
$m_{\bar{\ell}\nu}$ is the invariant mass of the lepton and neutrino, 
$\Gamma_t$ is the width of the top quark, 
$m_t$ and $m_b$ are the
masses of the top quark and b quark, respectively, and
$ E^{*}_{b} =  \frac{ m_t^2 - m_{b}^2 - m^2_{\bar{\ell}\nu}}{2 m_{\bar{\ell}\nu}} $.
The hadronic factor $\bar{F}_{had}$ is similar, with the exception that we do not distinguish between
up-type and down-type quarks from $W$-boson decay and use the average $\bar{F}_{had}$ related to the two permutations.
The background differential cross section uses the
sum of matrix elements for $W$+jets from the {\sc vecbos}~\cite{vecbos} Monte Carlo (MC) generator.

%%%%%%%%%%%%%%%%%%%%%%%%%%%%%%%%%%%%%%%%%%%%%%%%%%%%%%%%%%%%%%%%%%%%%%%%%%%%%%
%%%%%%%%%%%%%%%%% Detector  %%%%%%%%%%%%%%%%%%%%%%%%%%%%%%%%%%%%%%%%%%%%%%%%%%

The measurement is based on a data set with an integrated luminosity of 
approximately 2.7~$\rm{fb}^{-1}$
acquired by the Collider Detector at Fermilab (CDF II)~\cite{c:cdfDet} from $\ppbar$ 
collisions at $\sqrt{s}=1.96$~TeV.
The data used are collected using  
high-$p_T$ central (pseudorapidity~\cite{met} $|\eta|<1.1$) electron and muon triggers, 
a high-$p_T$ forward ($1.2<|\eta|<2.0$) electron trigger, 
and a trigger that requires large missing transverse 
energy $\met$~\cite{met} with either an energetic
electromagnetic cluster or two separated jets ($\met$+jets)~\cite{trigger}.
The $\met$+jets trigger is used to select additional events with high-$p_T$
muons, which are not selected by the lepton triggers.

%%%%%%%%%%%%%%%%%%%%%%%%%%%%%%%%%%%%%%%%%%%%%%%%%%%%%%%%%%%%%%%%%%%%%%%%%%%%%%
%%%%%%%%%%%%%%%% EventSel %%%%%%%%%%%%%%%%%%%%%%%%%%%%%%%%%%%%%%%%%%%%%%%%%%%%

Candidate events for the lepton plus jets 
final state ($\ttbar \to W^+ b W^- \bar{b} \to \ell \nu b q \bar{q}' \bar{b}$)
are selected to have a single,
isolated electron or muon candidate with $E_T>20$~GeV,
large imbalance in transverse momentum in the event ($\met >20$~GeV) 
as expected from the undetectable
neutrino, and at least four  jets with $E_T>20$~GeV. Jets are
reconstructed using a cone algorithm with radius 
$\Delta R=0.4$ in $\eta--\phi$ space,
and their energies are corrected for non-uniformities
in the calorimeter response as a function of jet $\eta$, multiple
$p \bar{p}$ interactions, and the hadronic jet energy scale of the
calorimeter~\cite{jesnim}.  
Of these jets, we
require at least one to have originated from a $b$ quark by using an
algorithm that identifies a long-lived $b$ hadron through the presence
of a displaced vertex ($b$ tag)~\cite{secvtx}.  Backgrounds to the
$t\bar t$ signal arise from multi-jet QCD production (QCD), $W$-boson
production in association with jets ($W$+jets), and electroweak
backgrounds (EWK) composed of diboson ($WW$, $WZ$, $ZZ$) and single
top-quark production.  
The $W$+jets background includes $b$-flavor jets
($W$+hf) as well as light flavor jets incorrectly 
identified as $b$ jets ($W$+lf). 

A detailed description of the background estimation can be
found in Ref.~\cite{xsection}. 
Table~\ref{tab:events} shows the
expected sample composition assuming a $t\bar t$ cross section of 6.7~pb.
There are overlapping events between those collected
by high-$p_T$ lepton triggers and $\met$+jets trigger.
These overlapping events are accounted in 
the central $e$/$\mu$ and forward $e$ categories,
and are explicitly vetoed from $\met$+jets 
category.

\begin{table}[h]
\caption{Number of expected and observed events in $2.7$ fb$^{-1}$
assuming a $\ttbar$ cross section of 6.7~pb.}
\begin{ruledtabular}\begin{tabular}{lccc}
 Process  & Central     &  Forward &  $\met$+jets \\
          & $e$, $\mu$  &    $e$   &    $\mu$       \\ \hline
\vspace*{-0.3cm}\\
$\ttbar$       & 478 $\pm$ 66 & 58 $\pm$ 8 & 134 $\pm$ 19 \\
$W$+hf         & 71 $\pm$ 22 & 13 $\pm$ 9 & 19 $\pm$ 6    \\
$W$+lf         & 23 $\pm$ 6 & 5 $\pm$ 7 & 6 $\pm$ 2      \\
EWK             & 17 $\pm$ 10 & 3 $\pm$ 1 & 5 $\pm$ 3      \\
QCD            & 28 $\pm$ 22 & 46 $\pm$ 37 & 1 $\pm$ 1    \\
\hline
Total expected        & 616 $\pm$ 74 & 125 $\pm$ 40 & 165 $\pm$ 20 \\
\hline
Observed       & 650  & 136  & 178    \\
\end{tabular}
\end{ruledtabular}
\label{tab:events} \rm
\end{table}

\begin{figure}[htb]
\centering
\includegraphics[width=0.5\textwidth]{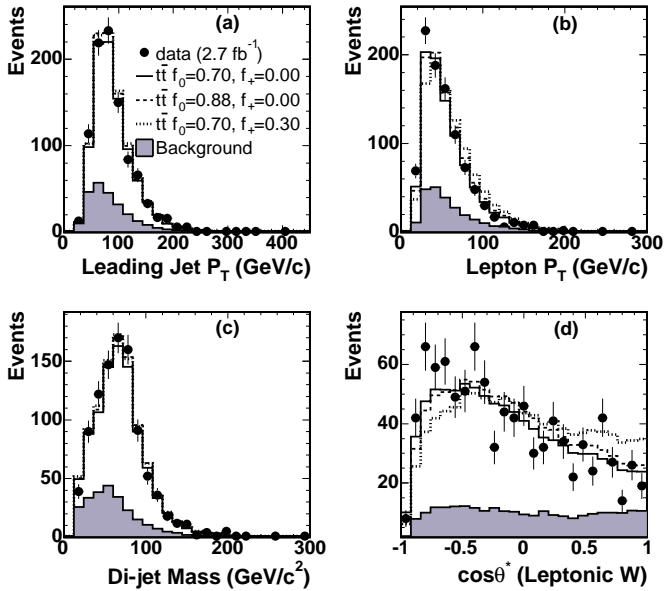}
\hfill
\caption{\label{fig:dataVal} Comparison of four kinematic variables for data and simulation
for different $W$ polarization fractions: solid, dashed and dotted histograms correspond to 
($f_0$, $f_+$) values of (0.7, 0.0), (0.88,0.0) and (0.7, 0.3), respectively. 
Plotted are (a) leading jet $p_T$, (b) lepton $p_T$; and for the reconstruction
chosen as most likely by the per-event likelihood
(c) the invariant mass of the pair of light quark jets from the hadronically 
decaying $W$ boson and (d) the $\Cts$ of the leptonically decaying $W$ boson.
}
\end{figure}

The {\sc herwig}~\cite{herwig} MC generator is used to model 
the $\ttbar$ signal events with $m_t = 175$ GeV/$\rm{c}^2$.
For estimation of various systematic uncertainties and background
modeling MC samples are created using the {\sc pythia}~\cite{pythia} generator, and
{\sc alpgen}~\cite{alpgen} or {\sc madevent}~\cite{Madevent} with {\sc pythia} or {\sc herwig} supplying the parton shower and fragmentation.
The QCD background is modeled using data control samples.
The signal and background modeling has been extensively checked.
Figure \ref{fig:dataVal} compares the observed data and the MC-predicted 
distributions of different kinematic variables.
We have validated the background model by studying a
high-statistics control sample of $W$+jets 
candidates extracted by vetoing events containing $b$-tagged jets.

%%%%%%%%%%%%%%%%%%%%%% Calibration  %%%%%%%%%%%%%%%%%%%%%%%%%%%%%%%%%
%%%%%%%%%%%%%%%%%%%%%%%%%%%%%%%%%%%%%%%%%%%%%%%%%%%%%%%%%%%%%%%%%%%%%

We calibrate the results of the likelihood fit using the simulated
$\ttbar$ and background samples, and the sample composition of table~\ref{tab:events}.
For the simultaneous measurement of $f_0$ and $f_+$, we find our
estimate $f_{0,\rm{m}}$ is related to the true value of $f_0$ by 
$f_{0,\rm{m}}= (0.88 \pm 0.02)  f_0 + (-0.12 \pm 0.01)$ and
our estimate of $f_{+,\rm{m}}$ is related to the true value of $f_+$ and $f_0$ by
$f_{+,\rm{m}}= (1.26 \pm 0.01)  f_+ + (0.17 \pm 0.02)  f_0 + (0.06 \pm 0.01)$.
We use these two equations and the measured polarization fractions to extract
the true polarization fractions. 
For our measurement
of $f_0$ with $f_+=0$, we find our estimate 
$f_{0,\rm{m}}= (1.15 \pm 0.04) f_0 + (-0.09 \pm 0.02)$,
and for our measurement
of $f_+$ with $f_0=0.7$, we find our estimate 
$f_{+,\rm{m}}= (1.17 \pm 0.05) f_+ + (0.01 \pm 0.01)$.
The uncertainties on the coefficients of the calibration functions
are accounted as the method-related systematic uncertainties, which 
cover possible biases due to the calibration procedure.
The differences between $f_{0,\rm{m}}$ or $f_{+,\rm{m}}$ values
and the corresponding true values are due to the fact that
the signal and background probabilities in the the likelihood do not model 
events with extra jets from initial and final state radiation (ISR/FSR), or
incorrect jet-parton assignment where a ISR/FSR jet is selected as one of the leading four jets,
or all the different background processes. 
Even though likelihood can be calculated only for the physical values of $f_0$ and $f_+$,
after calibration the corrected measured values can be slightly outside their physical ranges.

The robustness of the fitting procedure over all physical values of
($f_0$,$f_+$) has been tested with simulated experiments, using the number
of observed data events and the sample composition of table~\ref{tab:events}.  
In all cases, the fit is unbiased. Near the physical boundaries, we find that
the statistical uncertainty is underestimated by as much as a
factor of 1.5. We apply a correction to the statistical
uncertainty in these regions.
Assuming the SM, the expected statistical uncertainties
after all corrections
for the simultaneous measurement 
are $\pm0.116$ and $\pm0.074$ for $f_0$ and $f_+$, respectively.

%%%%%%%%%%%%%%%%%%%% Systematics %%%%%%%%%%%%%%%%%%%%%%
%%%%%%%%%%%%%%%%%%%%%%%%%%%%%%%%%%%%%%%%%%%%%%%%%%%%%%%

Various sources of systematic uncertainty affecting the measurement
are summarized in table \ref{tab:SystSummary}.
The leading sources of systematic uncertainty arise from 
MC modeling of initial and final state radiation (ISR/FSR), choice of 
parton distribution functions (PDFs),
choice of parton shower model, 
uncertainties on the measured jet energy, and 
the background shape and normalization.
The method-related uncertainty includes propagating the uncertainty 
on the fit parameters of the response curves, including their correlations.
All systematic uncertainties are determined by performing simulated experiments
in which the systematic parameter in question is varied,
the default method and calibrations are applied, and the shifts in the mean
measured polarization fractions are used to quantify the uncertainty.
All shifts are evaluated at the SM helicity fraction.

\noindent 
\begin{table}[h]
\caption[Summary of systematic uncertainties.]{Summary of systematic uncertainties.}
\begin{ruledtabular}\begin{tabular}{lcccc}
Source            & $\Delta f_0$  & $\Delta f_+$
                  & $\Delta f_0$  & $\Delta f_+$ \\ 
                  &               &     & \multicolumn{2}{c}{simultaneous} \\ \hline \hline
ISR/FSR           & 0.020  & 0.018  & 0.020  & 0.021 \\ 
PDF               & 0.024  & 0.013  & 0.009  & 0.016 \\ 
JES               & 0.018  & 0.017  & 0.004  & 0.012 \\ 
Parton shower     & 0.012  & 0.008  & 0.031  & 0.017 \\ 
Background        & 0.009  & 0.038  & 0.042  & 0.039 \\ 
Method-related    & 0.010  & 0.005  & 0.024  & 0.024 \\ 
b-tag SF          & 0.004  & 0.002  & 0.002  & 0.002 \\ \hline
Total             & 0.041  & 0.048  & 0.062  & 0.057 \\ 
\end{tabular}
\end{ruledtabular}
\label{tab:SystSummary}
\end{table}

%%%%%%%%%%%%%%%%%%%%%%%%%%%% Results %%%%%%%%%%%%%%%%%%%%%%%%%%
%%%%%%%%%%%%%%%%%%%%%%%%%%%%%%%%%%%%%%%%%%%%%%%%%%%%%%%%%%%%%%%

\begin{figure}[htb]
\centering
\includegraphics[width=0.35\textwidth]{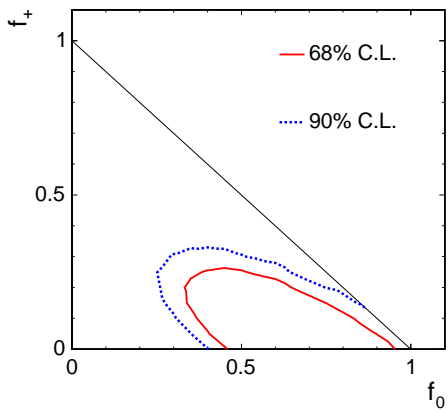}
\hfill
\caption{\label{fig:fig-test-CL-unCorrGrid-morePSE} 
Contours in the ($f_{0},f_{+}$) plane indicating 68\% and 90\% C.L. intervals
determined using Feldman-Cousins method. 
Note that the coverage is correct although
the center of the contours is not at the measured value obtained after
calibration.}
\end{figure}

For the simultaneous measurement of $f_0$ and $f_+$, we exclude the events
from the forward electron trigger as this significantly reduces the
systematic uncertainty from the background model. With 828 events and
after all corrections, we measure
\begin{eqnarray*}
   f_{0} & = & \:0.879 \pm 0.106 (\rm{stat}) \pm 0.062 (\rm{syst}) \\
   f_{+} & = &  -0.151 \pm 0.067 (\rm{stat}) \pm 0.057 (\rm{syst}).
\end{eqnarray*}
The statistical correlation between $f_0$ and $f_+$ is
$\rho=-0.59$. We estimate a
shift of $\mp(0.010 \pm 0.005)$ in $f_0$ and $\pm(0.017 \pm 0.003)$ in $f_+$
per $\pm1$ GeV/$\rm{c}^{2}$ shift in the top
quark mass from the central value of 175 GeV/$\rm{c}^{2}$.
As the central value is unphysical we have elected to ensure coverage by applying
the Feldman-Cousins method \cite{FC} to obtain the confidence level (C.L.) intervals shown in
Figure \ref{fig:fig-test-CL-unCorrGrid-morePSE}.

Fixing $f_+=0$ and with 964 events, we measure after all corrections
$f_{0} =  0.701 \pm 0.069 (\rm{stat}) \pm 0.041 (\rm{syst})$.
Fixing $f_0=0.70$, we measure after all corrections 
$f_{+} = -0.010 \pm 0.019 (\rm{stat}) \pm 0.049 (\rm{syst})$ and
find $f_+<0.12$ at 95\% C.L..  We estimate a shift of $\pm(0.011 \pm 0.003)$ 
in $f_0$ and $\pm(0.013 \pm 0.002)$ in
$f_+$ per $\pm1$ GeV/$\rm{c}^2$ shift in the top-quark mass from the central
value of 175 GeV/$\rm{c}^2$.

%%%%%%%%%%%%%%%%%%%%% Conclusion %%%%%%%%%%%%%%%%%%%%
%%%%%%%%%%%%%%%%%%%%%%%%%%%%%%%%%%%%%%%%%%%%%%%%%%%%%

In summary, we have measured the polarization of the $W$ boson in 
top-quark decays  using a matrix-element method in 2.7 $\rm{fb}^{-1}$ of CDF II data.
Our results are consistent with the SM. 
This result improves the combined statistical and systematic precision 
on both the model-independent and model-dependent determinations 
of the longitudinal polarization $f_0$ by a factor of 1.3
compared to the best previous measurement~\cite{CDFLatestWHelResults} 
for a 1.4 times increase in luminosity.

%%%%%%%%%%%%%%%%%%%%%%%%%%%%%%%%%%%%%%%%%%%%%%%%%%%%%%%%%%%
\newpage

%%%%%%%%%%%%%%%%%%%%%%%%%%%%%%%%%%%%%%%%%%%%%%%%%%%%%%%%%%%
%     Acknowledgements

We thank the Fermilab staff and the technical staffs of the participating institutions for their vital contributions. This work was supported by the U.S. Department of Energy and National Science Foundation; the Italian Istituto Nazionale di Fisica Nucleare; the Ministry of Education, Culture, Sports, Science and Technology of Japan; the Natural Sciences and Engineering Research Council of Canada; the National Science Council of the Republic of China; the Swiss National Science Foundation; the A.P. Sloan Foundation; the Bundesministerium f\"ur Bildung und Forschung, Germany; the World Class University Program, the National Research Foundation of Korea; the Science and Technology Facilities Council and the Royal Society, UK; the Institut National de Physique Nucleaire et Physique des Particules/CNRS; the Russian Foundation for Basic Research; the Ministerio de Ciencia e Innovaci\'{o}n, and Programa Consolider-Ingenio 2010, Spain; the Slovak R\&D Agency; and the Academy of Finland.

%%%%%%%%%%%%%%%%%%%%%%%%%%%%%%%%%%%%%%%%%%%%%%%%%%%%%%%%%
%     References
\bibliographystyle{unsrt}

\begin{thebibliography}{99}
\bibitem{ChargeConjugate} Charge-conjugate modes are included implicitly throughout this paper.
\bibitem{met}
We use a cylindrical coordinate system where the $z$ axis is
along the proton beam direction and $\theta$\ is the polar angle.
Pseudorapidity is $\eta = -\ln\tan(\theta/2)$, while transverse momentum is
$p_T = |p| \sin \theta$, and transverse energy is $E_T = E \sin \theta$.
Missing transverse energy, \met, is defined as the magnitude of $-\sum_iE_T^i \hat{n}_i$, where $\hat{n}_i$\ 
is the unit vector in the azimuthal
plane that points from the beam line to the $i^{\mathrm th}$ calorimeter
tower.
\end{thebibliography}

\end{document}